\documentclass{andp2012}
\usepackage[english]{babel}

\keywords{NEMS; Adiabatic; Transient dynamics; Out of equilibrium}
\title{Transient dynamics of an adiabatic NEMS}
\author[M. Biggio]{M. Biggio\inst{1}}
\author[F. Cavaliere]{F. Cavaliere\inst{2,3}\footnote{Corresponding author\quad E-mail:~\textsf{fabio.cavaliere@gmail.com}}}
\author[M. Storace]{M. Storace\inst{1}}
\author[M. Sassetti]{M. Sassetti\inst{2,3}}
\address[1]{DITEN, University of Genoa, Via Opera Pia 11a, 16145, Genova, Italy}
\address[2]{Dipartimento di Fisica, University of Genoa, Via Dodecaneso 33, 16146, Genova, Italy}
\address[3]{CNR-SPIN, Via Dodecaneso 33, 16146, Genova, Italy}
\shortauthors{M. Biggio et al.}

\begin{abstract}
This paper is focused on the transient dynamics of an adiabatic
nano-electromechanical system (NEMS), consisting of a nano-mechanical
oscillator coupled to a quantum dot. By numerically solving the
nonlinear stochastic differential equation governing the oscillator,
the time evolution of the oscillator position, of the dot occupation
number and of the current are studied. Different parameter settings
are studied where the system exhibits bi-stable, tri-stable or
mono-stable behavior on a finite-time horizon. It is shown that, after
a typically long transient, the system under investigation exhibits no
hysteretic behavior and that a unique steady state is reached,
independently of the initial conditions. The transient dynamics is
marked out by one or two well separated characteristic times,
depending on the considered case (i.e., mono- or multi-stable). We
evaluate these times for a dot on-resonance or off-resonance. It turns
out that the characteristic time scales are long in comparison to the
period of the uncoupled oscillator, particularly at low bias,
suggesting that the predicted transient dynamics may be observed in
state-of-the-art experimental setups.
\end{abstract}
\begin{document}
\maketitle

\section{Introduction}
Nanoelectromechanical systems (NEMS)~\cite{nemsrev1,nemsrev2}, consisting of a nanoscale oscillator coupled to the conduction electrons of a quantum dot, are intriguing systems from both an experimental and a theoretical point of view. Several different physical implementations of NEMS can be envisioned. The oscillator is commonly fabricated through suspended nano-beams~\cite{tang,roukes} or doubly-clamped carbon nanotubes (CNTs)~\cite{charlier,sapmaz,huttel,steele,gomez,cobdenhi,eronat}, whereas the quantum dot can be either nano-fabricated in a semiconducting host or embedded in the nanotube itself. {NEMS have attracted a considerable interest due to a wealth of possible applications, ranging from single-molecule mass spectrometers~\cite{naikm1,naikm2} to nanoscale gas sensors~\cite{gas1,gas2} and biosensors~\cite{bio1}.}\\
\noindent Basically, two markedly different regimes are possible for NEMS, according to the ratio between the bare oscillator frequency $\Omega_{0}$ and the average tunneling rate $\Gamma_{0}$ of electrons flowing through the quantum dot.\\

\noindent In the {\em anti-adiabatic} case ($\Omega_{0}\gg\Gamma_{0}$), the oscillator motion is revealed by the presence of quantized vibronic sidebands in the conductance of the system~\cite{sapmaz,cobdenhi,eronat,noirapcom,suspended}. One of the hallmarks of this regime is the Franck-Condon blockade, a peculiar low-bias current suppression occurring when electrons and vibrons are strongly coupled which has been theoretically predicted~\cite{koch,flensberg,mitra,koch2,noiscr} and experimentally observed~\cite{sapmaz,eronat}. Also peculiar features of the shot noise of conduction electrons have been predicted~\cite{koch,koch2,haupt,noinjp} and the influence of quantum coherences on the oscillator dynamics has been investigated~\cite{coherent,eidelstein}, possibly leading to an effective intrinsic cooling even in the anti-adiabatic regime.\\

\noindent In the opposite {\em adiabatic} case ($\Omega_{0}\ll\Gamma_{0}$), which we investigate in this paper, the bare oscillator dynamics is very slow in comparison to the electronic one. As a result, the oscillator behaves classically~\cite{blencowe,mozyrsky,blanter,boubsi}, with the ultra-fast motion of electrons giving rise to both an effective non-linear deterministic force acting on the oscillator and an effective non-linear damping~\cite{pistolesi,pistolaba,brandes,atalaya,brandes2,nocera,nocera2,nocera3,nocera4,clerk}. Moreover, the fluctuating nature of electron transport in the quantum dot gives rise to a nonlinear stochastic force acting on the oscillator~\cite{mozyrsky,brandes,nocera}. This makes the system an ideal condensed-matter playground for the study of nonlinear oscillations~\cite{oscrev,bachtoldnl,bachtoldnl2,bachtoldso,roukes2} and determines a very rich dynamical scenario, subject of many theoretical studies. Owing to the nonlinear nature of the oscillator dynamics, phenomena such as hysteresis, multi-stability and switching have been studied~\cite{komnik,mitra2,galperin1,galperin2,galperin3}. {Multistability in particular is a very intriguing feature envisioning NEMS as novel data storage elements. The key idea is to encode information in the different available states, to store data by preparing the system in one of the available ground states, employing the electronic subsystem (e.g. suitably tuning a gate voltage) and to possibly read out the dot state at a later stage~\cite{pistolesi}. In this respect, a crucial parameter to asses the robustness of the information storage element is the average dwell time of the states, i.e. the average time it takes for the system to switch between two different configurations. In this respect, the evolution towards a steady state may even be as detrimental since the stochastic jumps between different states renders the data storage useless. This fact motivates a thorough study of the {\em transient} dynamics of an adiabatic NEMS, going beyond the large number of works which have on the other hand characterized its {\em steady state}~\cite{blencowe,mozyrsky,boubsi,pistolesi,nocera,clerk}. In addition, also the transient dynamics of a NEMS has been the subject of investigation~\cite{tahir,riwar,wilner,albrecht,albrecht2,goker,soller,pulse1,pulse2}.} Two closely related issues arise in this context, namely
\begin{enumerate}
	\item whether the system exhibits a unique steady state or not;
	\item how the steady state is approached throughout the transient dynamics.
\end{enumerate}

\noindent Both issues have been addressed theoretically and no unique answer has been provided so far. By employing time-dependent Hartree-Fock (TDHF) techniques when $\Omega_{0}\lesssim\Gamma_{0}$~\cite{riwar}, multi-configuration TDHF (MCTDHF) techniques when $\Omega_{0}\ll\Gamma_{0}$~\cite{wilner} or even MCTDHF supplemented by diagrammatic Monte Carlo techniques~\cite{albrecht}, the tendency towards a non-unique steady state, depending on the initial condition of the oscillator, has been pointed out. However, the above numerical analyses mainly focused on rather {\em short} time scales of just a few periods of the uncoupled mechanical oscillator~\cite{albrecht}. On the other hand, by employing the polaron tunneling approximation for $\Omega_{0}\gtrsim\Gamma_{0}$ (thus, out of the adiabatic regime)~\cite{albrecht2} or in the presence of superconducting leads~\cite{soller} no hysteretic behavior has been reported.\\
\noindent The results cited above evidence one of the main difficulties in studying the time evolution of NEMS: the presence of possibly very long time scales~\cite{albrecht} makes it difficult to actually reach the steady state by means of numerical techniques.

\noindent In this paper, we study the time-dependent evolution of the NEMS observables - oscillator position, dot level occupation and current - as a function of different system parameters. We focus on the adiabatic case and consider a strong coupling between the oscillator and the electrons, the most favorable condition to observe multi-stability~\cite{pistolesi}, {considering different possible experimental realizations and discussing their potential and limitations.} Our task is to understand how the system evolves to a steady state and whether the latter is unique or not. Describing the system by means of the Anderson-Holstein model, the ultra-fast electronic dynamics is traced out, thus obtaining an effective nonlinear Langevin equation for the oscillator~\cite{pistolesi,brandes} and a corresponding Fokker-Planck equation for the probability density of the oscillator states~\cite{pistolesi}. By applying suitable numerical techniques, we solve the time-dependent Langevin equation and obtain the {\em full dynamical evolution} towards the steady state of the system observables with an arbitrary initial condition.\\

\noindent Our main findings are the following.
\begin{enumerate}
	\item Recasting the Langevin equation in a Fokker-Planck form allows to {\em prove} that the system exhibits a {\em unique} steady state which {\em does not} depend on the initial conditions.
	\item When the system is in a multi-stable setting, the transient dynamics is characterized by {\em two} or {\em three} time scales. The shortest ones are related to the dynamical trapping of the oscillator state around one of the equilibrium positions of the system. The longest one represents the characteristic relaxation time towards the (unique) steady state.
	\item When the system is in a mono-stable setting, the transient dynamics is characterized by the relaxation time only.
\end{enumerate}
These characteristic times have been estimated by studying the statistical properties of the solutions of the Langevin equation and the spectral properties of the Fokker-Planck equation. Especially in the case of a multi-stable NEMS, they are found to be {\em several orders of magnitude larger} than the typical oscillator period $1/\Omega_{0}$, supporting some of the claims made by other authors~\cite{albrecht,albrecht2}. We remark that our numerical approach explores the system up to very long time scales, allowing to observe the fully developed steady state. {In the multistable case, the shortest time scale is identified with the dwell time, which ultimately limits the performance of an hypothetical NEMS data storage element.} The characteristic time scales are found to be within the reach of state-of-the-art experimental investigations for many systems of interest. {In particular, suspended CNTs seem ideal candidates in view of their ability to reach the strong coupling adiabatic regime.}\\

\noindent The outline of the paper is as follows. In Sec.~\ref{sec:sec1} the model of the system and the methods employed are presented. In Sec.~\ref{sec:sec2} the results are presented and discussed, {including possible realistic implementations of the NEMS under investigation}. Some conclusions are drawn in Sec.~\ref{sec:sec3}.

\section{Model and Methods}
\label{sec:sec1}
The NEMS is modeled as a single-level quantum dot linearly coupled to a harmonic vibrational mode. The dot and oscillator Hamiltonians are ($\hbar=1$)
\begin{eqnarray}
H_{d}&=&\epsilon d^{\dagger}d\, ,\label{eq:h0d}\\
H_{o}&=&\frac{P^2}{2m}+\frac{m\Omega_{0}^2}{2}X^2.
\label{eq:h0v}
\end{eqnarray}
Here, $\epsilon$ is the dot level position (which can be conveniently tuned via a capacitively coupled gate) and $d$ is the dot fermionic annihilation operator. We assume here to deal with a dot characterized by a large charging energy, so that double occupancy can be neglected. Also, to keep the notation simple, a spinless model has been employed~\cite{pistolesi}. The extension of our results to the spinful case is straightforward and does not lead to qualitatively different conclusions. Position $X$ and momentum $P$ operators for the oscillator (with mass $m$) have been introduced, being $\Omega_{0}$ the bare oscillator frequency. The dot and the oscillator are coupled by the term
\begin{equation}
H_{d-o}=\lambda X d^{\dagger}d\, ,\label{eq:hdv}
\end{equation}
where $\lambda$ is the coupling force between electrons and vibrations. {An example of the explicit form of $\lambda$ for the relevant case of a suspended CNT is provided in Sec.~\ref{sec:obs}}. The dot is also coupled to left and right contacts of free electrons described by
\begin{equation}
H_{l}=\sum_{\alpha=L,R}\sum_{k}\varepsilon_{\alpha,k} c^{\dagger}_{\alpha,k}c_{\alpha,k}\, .\label{eq:h0l}
\end{equation}
$c_{\alpha,k}$ is the Fermi operator of lead $\alpha=L,R$ and $k$ is the momentum of electrons therein. Contacts are kept at a given electrochemical potential $\mu_{\alpha}=\mu_{0}+eV_{\alpha}$ where $e$ is the electron charge, $V_{\alpha}$ a bias voltage applied to lead $\alpha$ and $\mu_{0}$ the equilibrium chemical potential, assumed to be equal in both leads in the absence of bias. Henceforth we assume a symmetrical biasing with $V_{L}=-V/2$ and $V_{R}=V/2$. Dot and contacts are tunnel-coupled, with the tunneling Hamiltonian
\begin{equation}
H_t=\sum_{\alpha=L,R}\sum_{k}\chi_{\alpha}d^{\dagger}c_{\alpha,k}+\mathrm{h.c.}\, .\label{eq:ht}
\end{equation}
The above Hamiltonians constitute the Anderson-Holstein model. {This deceptively simple model produces a very rich and interesting physics and has been successfully employed to describe NEMS in a vast range of regimes~\cite{koch,flensberg,mitra,pistolesi,brandes,mozyrsky,galperin1,riwar}}. The dot-oscillator coupling sets the characteristic polaron energy and length scales
\begin{equation}
E_{p}=\frac{\lambda^2}{2m\Omega_{0}^2}\ ;\ \ell_{p}=\frac{\lambda}{m\Omega_{0}^2}\, ,\label{eq:enlensc}
\end{equation}
whereas the tunnel coupling sets the average tunneling rate $\Gamma_0=2\pi|\chi|^2$ where, for simplicity, we assume symmetric tunnel barriers $\chi_\alpha=\chi$ ($\alpha=L,R$).\\
\noindent In the {\em adiabatic} assumption ($\Omega_{0}\ll\Gamma_{0}$), the vibrational dynamics follows adiabatically the ultra-fast motion of electrons, which act as an effective force and a dissipative bath on the oscillator~\cite{mozyrsky,brandes,pistolesi,nocera,grifoni}. Standard path-integral techniques~\cite{weiss,grifoni,mozyrsky,brandes} allow to trace out the degrees of freedom of electrons in both the contacts and the quantum dot, thus leading to an effective Langevin equation of motion for the oscillator
\begin{equation}
\ddot{x}+A(x)\dot{x}-F(x)=\sqrt{D(x)}\xi(\tau)\, ,\label{eq:lan}	
\end{equation}
where $x=X/\ell_{p}$ is the dimensionless oscillator position (regarded as a classical variable) and overdots imply derivatives w.r.t. $\tau=\Omega_0 t$, the dimensionless time variable. The dimensionless Eq.~(\ref{eq:lan}) is characterized by a {\em non-linear and positive definite friction coefficient} $A(x)$, an effective potential $U(x)$ with $F(x)=-\partial_{x}U(x)$ and a multiplicative noise term $\sqrt{D(x)}\xi(\tau)$.\\
\noindent The nonlinear force term $F(x)$ is due to the coupling of the oscillator to the dot and arises at the lowest (zero-th) order in $\Omega_{0}/\Gamma_{0}$ (i.e., in the limit $m\to\infty$)~\cite{nocera}. The nonlinear damping and noise terms, on the other hand, represent the first-order contributions in $\Omega_{0}/\Gamma_{0}$ to the oscillator dynamics~\cite{nocera} (adiabatic corrections) and are related via the fluctuation-dissipation theorem~\cite{brandes}. In particular, the stochastic nature of the ultra-fast electron motion gives rise to random fluctuations of the dot occupation. These in turn are represented by the stochastic forcing term $\sqrt{D(x)}\xi(t)$, where $\xi(\tau)$ is a white Gaussian noise~\cite{mozyrsky} with $\langle\xi(\tau)\rangle=0$ and $\langle\langle \xi(\tau)\xi(\tau')\rangle\rangle=\delta(\tau-\tau')$. Here $\langle\langle\ldots\rangle\rangle$ denotes an average over the realizations of the stochastic process $\xi(\tau)$.\\
\noindent{In this work, we do not consider explicitly {\em extrinsic} damping mechanisms due to possible external damping baths. Their expected impact on the results which we will present below are briefly addressed in Sec.~\ref{sec:validity}.}

\noindent The dimensionless damping, force and noise terms are expressed by means of the forward and backward dot Green's functions on the Keldysh contour $G_{\pm}(\omega,x)$ as~\cite{pistolesi}
\begin{eqnarray}
A(x)&=&\frac{\omega}{2\pi}\int\mathrm{d}\omega'\ G_{+}(\omega',x)\partial_{\omega'}G_{-}(\omega',x)\, ,\label{eq:Ag}\\	
F(x)&=&-x+\frac{i}{2\pi}\int\mathrm{d}\omega'\ G_{+}(\omega',x)\, ,\label{eq:Fg}\\	
D(x)&=&\frac{\omega}{2\pi}\int\mathrm{d}\omega'\ G_{+}(\omega',x)G_{-}(\omega',x)\, ,\label{eq:Dg}		
\end{eqnarray}
where $\omega=\Omega_{0}/(2E_{p})$. At small temperatures ($k_{B}T\ll\Gamma_{0}$), the Green's functions are~\cite{pistolesi}
\begin{equation}
G_{\pm}(\omega,x)=\pm i\gamma\sum_{\alpha=L,R}\frac{\theta\left(\pm\frac{\mu_{\alpha}}{2E_{p}}\mp\omega\right)}{\left(\omega-\varepsilon-x\right)^2+\gamma^2}\, ,\label{eq:gf}
\end{equation}
with $\gamma=\Gamma_{0}/(2E_{p})$, $\varepsilon=\epsilon/2E_{p}$, and $\theta(x)$ the Heaviside step function. By substituting Eq.~(\ref{eq:gf}) into Eqns.~(\ref{eq:Ag}-\ref{eq:Dg}) one obtains~\cite{nocera,pistolesi}
\begin{eqnarray}
A(x)&=&\frac{2\omega}{\pi\gamma^2}\sum_{\alpha=L,R}\frac{1}{\left(1+\Delta_{\alpha}^2\right)^2} \, ,\label{eq:A}\\	
U(x)&=&\frac{x(x+1)}{2}-\frac{\gamma}{4\pi}\sum_{\alpha=L,R}\left[\Delta_{\alpha}\arctan\left(\Delta_{\alpha}\right)\right.\nonumber\\
&&+\left.\log\left(\frac{1}{\sqrt{1+\Delta_{\alpha}^{2}}}\right)\right]\, ,\label{eq:F}\\	 
D(x)&=&\frac{\omega}{\pi\gamma}\sum_{\alpha=L,R}s_{\alpha}\left[\arctan\left(\Delta_{\alpha}\right)+\frac{\Delta_{\alpha}}{1+\Delta_{\alpha}^2}\right] \, ,\label{eq:D}\\	
\end{eqnarray}
where $s_{L}=+1$, $s_{R}=-1$ and
\begin{eqnarray}
\Delta_{L}&=&\frac{u+2(\varepsilon-x)}{\gamma} \, ,\label{eq:dl}\\
\Delta_{R}&=&-\frac{u-2(\varepsilon-x)}{\gamma} \, ,\label{eq:dr}
\end{eqnarray}
with $u=|e|V/(2E_{p})$. The shape of the potential energy $U(x)$ (and then the oscillator dynamics) changes according to $\varepsilon$ (related to the normalized gate potential) and $u$ (the normalized voltage bias between the leads), which are chosen as bifurcation parameters.\\

\noindent Equation~(\ref{eq:lan}) is a stochastic differential equation (SDE). Therefore, given the $\nu$-th realization of the noise process $\xi_\nu(\tau)$, the solution $x_{\nu}(\tau)$, subject to initial conditions $x_{\nu}(0)=x_0$ and $\dot{x}_{\nu}(0)=v_0$, fluctuates stochastically. A {\em probability density} $\mathcal{P}(x,v;\tau)$ for the oscillator to occupy the state $(x,v)$ (with $v=\dot{x}$) at normalized time $\tau$ can be introduced as
\begin{equation}
\mathcal{P}(x,v;\tau)=\langle\langle\delta\left(x-x_{\nu}(\tau)\right)\delta\left(v-\dot{x}_{\nu}(\tau)\right)\rangle\rangle\, .\label{eq:prob}
\end{equation}
The probability density $\mathcal{P}(x,v;\tau)$ is obtained numerically by solving the SDE for a fairly large number of different realizations of the stochastic process $\xi_{\nu}(\tau)$ by means of a highly optimized parallel algorithm. Solution traces for $x(t)$ and $v(t)$ are then sampled and a histogram is created for $\mathcal{P}(x,v;\tau)$. The process is iterated until convergence on the probability density is achieved. We remark that this procedure is {\em not} restricted to the asymptotic ($\tau\to\infty$) case. Indeed, as already anticipated, we are particularly interested into the transient evolution of the NEMS.\\
\noindent An alternative approach to obtain $\mathcal{P}(x,v;\tau)$ can be pursued, casting the Langevin equation (\ref{eq:lan}) into a Fokker-Planck equation~\cite{risken} $\dot{\mathcal{P}}(x,v;\tau)=\mathcal{L}\left[\mathcal{P}(x,v;\tau)\right]$, where the operator $\mathcal{L}$ is given by~\cite{pistolesi}
\begin{equation}
\mathcal{L}\left[{\mathcal P}\right]=-v\frac{\partial{\mathcal P}}{\partial x}-F(x)\frac{\partial{\mathcal P}}{\partial v}+A(x)\frac{\partial}{\partial v}\left(v{\mathcal P} \right)+D(x)\frac{\partial^{2}{\mathcal P}}{\partial v^2}\, .\label{eq:liouv}
\end{equation}
Both methods are employed here, since they bring complementary information on the system behavior.\\

\noindent Once $\mathcal{P}(x,v;\tau)$ is obtained, the expectation value of any given observable $\mathcal{O}(x,v)$ at time $\tau$ can therefore be evaluated as $\langle\mathcal{O}(\tau)\rangle=\int\mathrm{d}x\ \int\mathrm{d}v\ \mathcal{P}(x,v;\tau)\mathcal{O}(x,v)$. When these quantities do not depend on $v$ but only on $x$, these expression simplify to $P(x;\tau)=\int\mathrm{d}v\ \mathcal{P}(x,v;\tau)$ and $\langle \mathcal{O}(\tau)\rangle=\int\mathrm{d}x\ P(x;\tau)\mathcal{O}(x)$.\\ 
\noindent In this paper, we focus on three quantities, namely the position $x$, the dot current and its occupation number, given for $k_{B}T\ll\Gamma_{0}$ by~\cite{nocera}
\begin{eqnarray}
I(x)&=&\frac{1}{2\pi}\sum_{\alpha=L,R}s_{\alpha}\arctan\left(\Delta_{\alpha}\right)\, \label{eq:I}\\
n(x)&=&\frac{1}{2}+\frac{1}{2\pi}\sum_{\alpha=L,R}\arctan\left(\Delta_{\alpha}\right) \, .\label{eq:n}
\end{eqnarray}
The latter quantities depend parametrically on $x$ in the adiabatic regime~\cite{nocera,pistolesi}, while the time dependence of their averages is encoded in $\mathcal{P}(x;\tau)$.\\

\section{Results}
\label{sec:sec2}
\subsection{Effective potential landscapes}\label{subsec:subsec2}
{In the following, we will consider the strong-coupling regime~\cite{pistolesi}
\begin{equation}
\gamma<1	
\end{equation}
which, together with the adiabatic hypothesis $\omega\ll\gamma$ to satisfy the adiabatic condition), restricts $\omega\ll 1$. As we will see, this is the most favorable regime to observe multi-stability and switching phenomena~\cite{pistolesi}. From the physical point of view, the strong coupling regime corresponds to setting $E_{p}$ as the largest energy scale.}\\
\noindent The equilibrium conditions for the noiseless system (Eq.~\ref{eq:lan} with $D(x) = 0$) are $v=0$ and $F(x) = -\partial_{x}U(x) = 0$. Then, the equilibrium points are the extrema of $U(x)$.
\begin{figure}[htb!]
 \centering
 \includegraphics[width=\columnwidth,keepaspectratio]{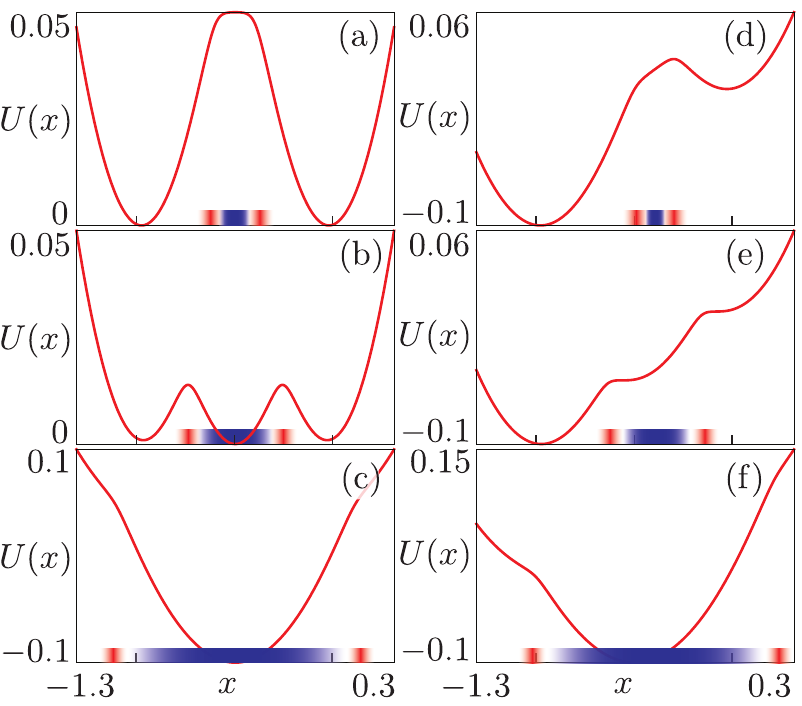}
 \caption{\label{fig:fig1} Effective potential $U(x)$ for different values of $\varepsilon$ and $u$. Left column: on-resonance case, $\varepsilon = -0.5$. Right column: off-resonance case, $\varepsilon = -0.4$. (a,d) $u=0.2$; (b,e) $u=0.475$; (c,f) $u=1.0$. The red (blue) shade marks the regions where $A(x)$ ($D(x)$) in Eqns.~(\ref{eq:A},\ref{eq:D}) is maximal. Other parameters: $\omega=10^{-3}$, $\gamma=0.08$.}
\end{figure}
We start discussing the shape of $U(x)$, by varying the parameters $u$ and $\varepsilon$. The effective potential $U(x)$ is shown in Fig.~\ref{fig:fig1} for an on-resonance ($\varepsilon=-0.5$~\cite{pistolaba}, panels a-c) and off-resonance ($\varepsilon=-0.4$, panels d-f) dot. At low $u$ values, two minima appear, around $x=0$ and $x=-1$, for both the on-resonance (panel a) and the off-resonance (panel d) cases. Increasing $u$, the on-resonance case (panel b) is characterized by three minima - the two discussed above and a new one developing at $x=-0.5$. On the other hand, the off-resonance effective potential can have either two or three minima. Panel e displays a limit case with two minima and a third ``ghost'' minimum (see also Fig.~\ref{fig:fig2}b).\\
\noindent For $u=1$ or larger, only the minimum at $x\approx-0.5$ survives for both the on-resonance and off-resonance cases (panels c and f).\\
\noindent {To illustrate the importance of the strong coupling regime in shaping $U(x)$ and giving rise to multistability, let us consider the most favorable case, that of a NEMS in resonance $\varepsilon=-0.5$ in the low bias regime $u\ll 1$. Here, the potential barrier $\Delta U=U(-1/2)-U(0)=U(-1/2)-U(-1)$, separating the two degenerate minima, can be estimated as
\begin{equation}
\Delta U=\frac{1}{2\pi}\left\{\arctan\left(\frac{1}{\gamma}\right)+\gamma\log\left[\frac{\gamma}{\sqrt{1+\gamma^2}}\right]\right\}-\frac{1}{8}\, .	
\end{equation}
One finds that for $\gamma\lesssim 1$ one has $\Delta U>0$, while increasing $\gamma$ above that threshold results in a progressive reduction of $\Delta U$ and the ultimate disappearance of bistability.}\\

\begin{figure}[htb!]
 \centering
 \includegraphics[width=\columnwidth,keepaspectratio]{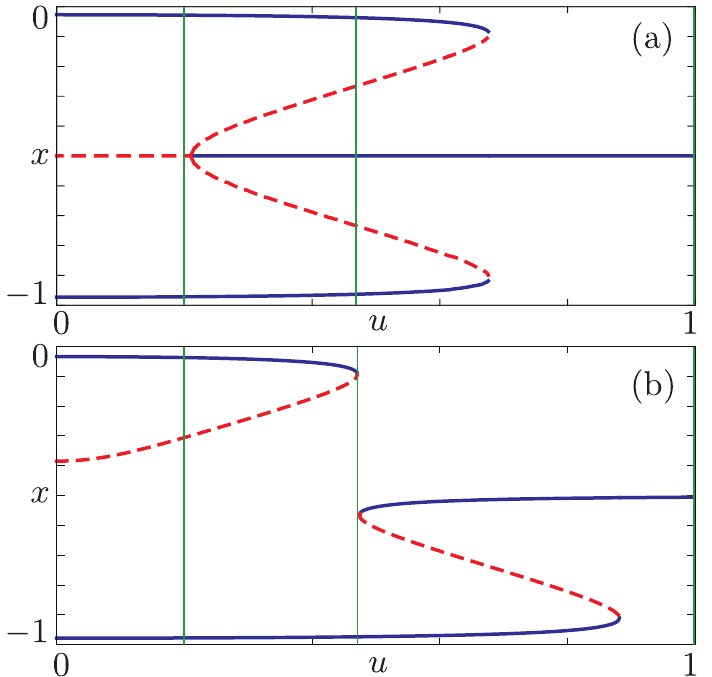}
 \caption{\label{fig:fig2}Position of stable (solid line) and unstable (dashed line) equilibrium positions for the system as a function of the normalized bias $u$ for (a) $\varepsilon=-0.5$ and (b) $\varepsilon=-0.4$. The green lines represent the values of $u$ at which the plots in Fig.~\ref{fig:fig1} have been made.}
\end{figure}
Figure~\ref{fig:fig2} shows the position of the minima (represented by a solid line, and corresponding to stable equilibria for the noiseless NEMS) and maxima (represented by a dashed line, and corresponding to unstable equilibria) of $U(x)$ as a function of $u$. For low values of $u$, in both the on-resonance (panel a) and off-resonance (panel b) cases, the presence of a {\em bi-stable} region is evident. As $u$ increases, in panel a a subcritical pitchfork bifurcation occurs and the system enters a tri-stable region. Two fold bifurcations at still higher $u$ values mark the end of the tri-stable region and the beginning of a mono-stable region with only the minimum located at $x=-0.5$. Tuning the system off-resonance by changing $\varepsilon$ induces a symmetry breaking in the diagram described above. When moving far enough from resonance, the tri-stable region even disappears. Indeed, panel b shows the limit case ($\varepsilon=-0.4$) where the tri-stable region disappears.\\
\noindent There is a close relationship between the oscillator and the dot states. In particular, for given $u$ and $\varepsilon$, each minimum of $U(x)$ corresponds to a different behavior of the quantum dot. The two minima at $x\approx 0$ and $x\approx -1$, occurring for low to intermediate $u$ values, correspond to situations where the dot is essentially locked into either the empty ($n=0$) or the occupied ($n=1$) charge state respectively. To exemplify this, let us consider the low-bias case ($u\approx 0$), where
\begin{eqnarray}
n(0)&\approx&\frac{1}{2}+\frac{1}{\pi}\arctan\left(\frac{2\varepsilon}{\gamma}\right)\, ,\\
n(-1)&\approx&\frac{1}{2}+\frac{1}{\pi}\arctan\left[\frac{2\left(\varepsilon+1\right)}{\gamma}\right]\, .
\end{eqnarray}
For $\varepsilon>-1$ and $\gamma\ll 1$, one has $n(0)\approx 0$ and $n(-1)\approx 1$. This situation is the semiclassical counterpart of the well known Coulomb blockade regime~\cite{pistolesi}.\\
\noindent The third stable minimum only arises for $u\geq u^{*}(\varepsilon)$. For instance, for $\varepsilon=-0.5$ one finds $u^{*}(-0.5)\approx\sqrt{2\gamma/\pi}$. In general, as exemplified in Fig.~\ref{fig:fig2}(b), one finds that $u^{*}(\varepsilon\neq-0.5)>u^{*}(-0.5)$. An approximation for the position of this minimum is
\begin{equation}
x^{*}(\varepsilon)=-\frac{1}{2}-\frac{\gamma\left(2\varepsilon+1\right)}{\pi\left(u^2+\gamma^2\right)-2\gamma}
\end{equation}
valid for $u\geq u^{*}(\varepsilon)$ and $|\varepsilon+1/2|\lesssim\gamma$. Roughly speaking, for $u\gtrsim 1$, $x^{*}\approx -0.5$. Additionally, it can be checked numerically that $n\left[x^{*}(\varepsilon)\right]\approx 0.5$ (with $n\left[x^{*}(-0.5)\right]\equiv 0.5$) and that for the same values the current exhibits a maximum. This situation is hence closely reminiscent of the sequential tunneling regime, where for symmetric tunnel barrier one expects half dot filling and maximal current through the system at finite bias~\cite{pistolesi,pistolaba}.\\
\noindent What described above holds for the noiseless system. The presence of the electronic noise induces crucial modifications in the above simple picture. In particular, it induces {\em jumps} between the different minima of $U(x)$ in the bi-stable and tri-stable cases, and stochastic fluctuations around the minimum of $U(x)$ in the mono-stable case. The regions where $D(x)$ is largest are marked with a blue shade in Fig.~\ref{fig:fig1}, whereas the regions where $A(x)$ is maximal are marked by red shades.\\
\noindent Due to these jumps, the oscillator cannot remain indefinitely stuck into one of the potential wells. This already suggests that the system may reach a {\em dynamical} steady state characterized by stochastic jumps between minima while the occupation probability $\mathcal{P}(x,v;\tau)$ attains a steady shape independent of the particular initial condition. This fact will be proven in the next section.
\subsection{Uniqueness of the steady state}
\label{subsec:uniqueness}
The adiabatic approximation considered here leads to the Fokker-Planck operator Eq.~(\ref{eq:liouv}) defines a {\em steady Fokker-Planck equation}. For this equation, a {\em unique steady state} exists, for any initial condition~\cite{unique}. To prove this fact, let us introduce the quantities
\begin{equation}
\mathbf{y}=\left(\begin{array}{c}v\\x\end{array}\right)\ ;\ \mathbf{a}=\left(\begin{array}{c}F(x)-vA(x)\\v\end{array}\right)\ ;\ \mathbf{d}=\left(\begin{array}{cc}D(x) & 0\\ 0 & 0\end{array}\right)\, ,\label{eq:matd}
\end{equation}
which allow recasting the Fokker-Planck operator as follows
\begin{equation}
\mathcal{L}=-\frac{\partial}{\partial y_{\mu}}\left[a_{\mu}\cdot\ \right]+\frac{1}{2}\frac{\partial^2}{\partial y_{\mu}\partial y_{\nu}}\left[d_{\mu,\nu}\cdot\ \right]
\end{equation}
where $\mu,\nu=1,2$ and a summation over repeated indices is implied. In order for the Fokker-Planck equation to be steady and exhibit an unique solution, two conditions have to be met~\cite{unique}. First of all, the matrix $\mathbf{d}$ must contain a positive-definite sub-matrix, which is obvious since $D(x)>0$. The second condition is that the system of partial differential equations
\begin{eqnarray}
\frac{\partial g(x,v;\tau)}{\partial x}&=&0\label{eq:pde1}\\
\frac{\partial g(x,v;\tau)}{\partial \tau}&=&-\left[F(x)-vA(x)\right]\frac{\partial g(x,v;\tau)}{\partial v}\label{eq:pde3}
\end{eqnarray}
has the unique solution $g(x,v;\tau)\equiv\mathrm{const.}$~\cite{unique}. This can be easily proven by further deriving Eq.~(\ref{eq:pde3}) with respect to $x$ and taking into account Eq.~(\ref{eq:pde3}), which implies
\begin{equation}
\left[\partial_{x}F(x)-v\partial_{x}A(x)\right]\frac{\partial g(x,v;\tau)}{\partial v}=0\, .
\end{equation}
Since the quantity within square brackets is in general non-vanishing, one can conclude that $\partial_{v}g(x,v;\tau)=0$, which in turn implies (owing to Eq.~(\ref{eq:pde3})) $\partial_{\tau}g(x,v;\tau)=0$ and shows that $g(x,v;\tau)$ is constant.\\
\noindent This proves that our Fokker-Planck is steady, from which stems that there is a {\em unique} steady state {\em regardless the initial condition} of the system~\cite{unique}. Therefore, any hysteretic behavior can be ruled out, in contrast to previous results obtained by means of mean field methods~\cite{riwar,wilner,albrecht}. We want to stress here that the nature of this steady state is {\em dynamical}: indeed, when the steady state is reached one has that $\mathcal{P}(x,v;\tau)$ is independent of $\tau$ but $x(\tau)$ and $v(tau)$ stochastically fluctuate due to the electronic noise term in the Langevin equation.\\
\noindent Having proven that the steady state is unique, we still have to understand {\em how} the system approaches this unique steady state studying the full time evolution of the oscillator and dot variables, which constitutes the main task of this paper.
\subsection{Convergence to the steady state}
For a given set of dot and oscillator parameters, we have solved Eq.~(\ref{eq:lan}) starting from different initial conditions for the oscillator and following the evolution of oscillator position $x$, dot occupation $n(x)$ and current $I(x)$ (averaged over the noise realizations), until a steady state is reached. 

\subsubsection{On-resonance case ($\varepsilon=-0.5$)}
\begin{figure}[htb!]
 \centering
 \includegraphics[width=\columnwidth,keepaspectratio]{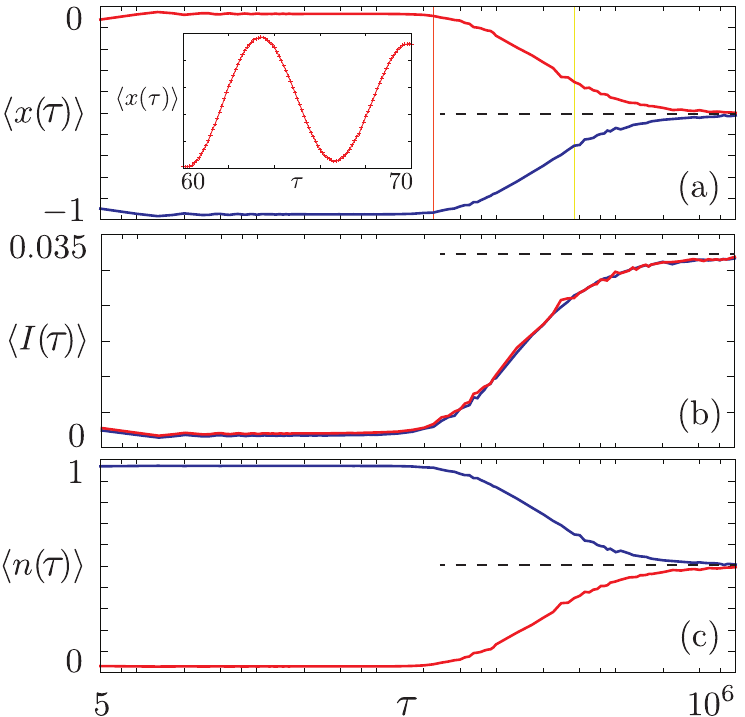}
 \caption{\label{fig:fig3} (a) $\langle x(\tau)\rangle$; (b) $\langle I(\tau)\rangle$; (c) $\langle n(\tau)\rangle$ for $u=0.2$, $\varepsilon=-0.5$ and different initial conditions $x_0=0, v_0=0$ (red curve) and $x_0=-1, v_0=0$ (blue curve). The inset in panel a shows a zoom of the short-time behavior of $\langle x(\tau)\rangle$ for the initial condition $x_0=0, v_0=0$. The orange (yellow) vertical line marks the characteristic dwell (relaxation) time scale $\tau_{d}$ ($\tau_{r}$) (see text for details). The steady state values are denoted by a dashed line. Other parameters: $\omega=10^{-3}$, $\gamma=0.08$.}
\end{figure}

Figure~\ref{fig:fig3} shows plots of $\langle x(\tau)\rangle$, $\langle I(\tau)\rangle$ and $\langle n(\tau)\rangle$ for two different initial conditions in the bi-stable region with bias $u=0.2$. 
All curves converge in the very long time limit to a common steady state. We stress that, in accordance to what has been proven in Sec.~\ref{subsec:uniqueness}, there is convergence to a unique steady state in {\em all} cases that we have investigated. The two initial conditions chosen here set the oscillator at rest near either of the minima of $U(x)$ for $\tau=0$.\\
\noindent The dynamics shows two very distinct phases. Initially, the
oscillator evolution is essentially {\em frozen} within the potential
wells. Indeed, a close-up of this region, shown in the inset of panel
a for $\langle x(\tau)\rangle$, displays a damped oscillatory behavior
with a quasi-period given by the bare oscillator frequency. In this
situation the dynamics is ruled by a competition between friction and
noise (see Fig.~\ref{fig:fig1}(a)). Indeed, the damped oscillations
are due to the weak tails of the nonlinear damping $A(x)$. However,
since the noise term is even smaller than $A(x)$ near the minima of
$U(x)$, the oscillator can settle near the minima. While the
oscillator is locked near the potential minima, the dot is in a
well-defined average state, empty or full, depending on the occupied
well. This agrees with the discussion in Section \ref{subsec:subsec2}
and shows how preparing the oscillator in a certain state
correspondingly locks the dot charge. Only after a certain time, the
small noise terms are able to induce transitions between the two
minima. This marks the onset of a new system dynamics, dominated by
{\em stochastic jumps between the potential minima, induced by the
  noise}. As soon as this transition sets in, the system begins to
evolve towards the unique steady state, which is eventually reached
within a given time scale. Notice that the time scale needed to reach
the steady state does not depend on the initial condition of the
oscillator. In the steady state, the current is larger than in the
transient evolution, due to the (partial) occupation of the region
around $x=-0.5$, where the current reaches its maximum value. It has
to be pointed out that the envelopes of $\langle x(\tau)\rangle$,
$\langle I(\tau)\rangle$, $\langle n(\tau)\rangle$ cannot be simply
fitted by a simple exponential function (not shown), confirming the multiple
time-scale dynamics of the system.\\
\noindent We now define the two relevant time scales which have been described above and which characterize the transient dynamics towards the steady state. We define {\em dwell time} $\tau_{d}$ the average time that the system spends in the "frozen" configuration, and {\em relaxation time} $\tau_{r}$ the typical time scale over which the system reaches the steady state. Of course, $\tau_{d}<\tau_{r}$. As we will see later, $\tau_{d}$ may depend on the initial condition of the system and even collapse to zero in certain situations. On the other hand, $\tau_{r}$ is always different from zero and essentially independent of the initial conditions.\\
\begin{figure}[htb!]
 \centering
 \includegraphics[width=\columnwidth,keepaspectratio]{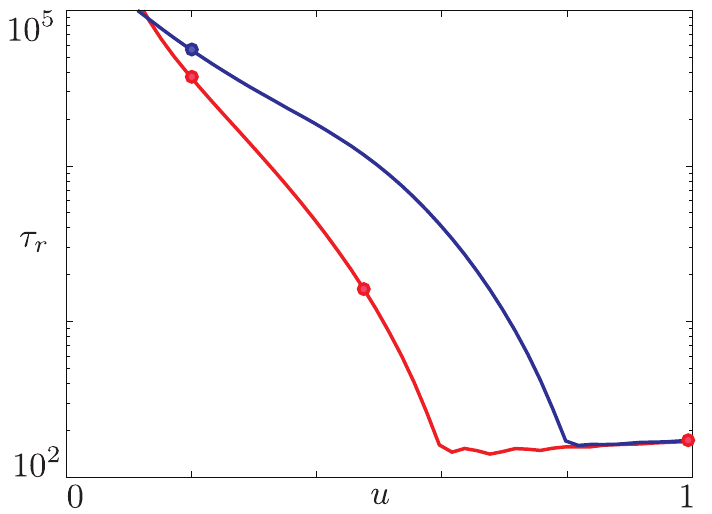}
 \caption{\label{fig:fig4} Relaxation time $\tau_{r}$ (see text) as a function of $u$ for $\varepsilon=-0.5$ (red curve) and $\varepsilon=-0.4$ (blue curve). The dots correspond to the cases shown in Figs.~\ref{fig:fig3},~\ref{fig:fig3},~\ref{fig:fig7},~\ref{fig:fig8},~\ref{fig:fig9}. Other parameters: $\omega=10^{-3}$, $\gamma=0.08$.}
\end{figure}
\noindent No multi-stability is hence found. However, in the case discussed above both $\tau_{d}$ and $\tau_{r}$ are very long in comparison to the bare period of the uncoupled oscillator. This suggests that the relaxation towards a steady state can only be observed if an experiment (or a calculation) is performed up to very long times in comparison to the bare oscillator period.\\
\noindent The two time scales are obtained analyzing the long-time behavior of the solutions $x(\tau)$ of Eq.~(\ref{eq:lan}) and of the operator $\mathcal{L}$ in Eq.~(\ref{eq:liouv}).\\
\noindent The relaxation time scale $\tau_{r}$ is inferred by a numerical analysis of the Fokker-Planck equation. Indeed, the existence of a unique steady state ensures the presence of only one zero eigenvalue, $e_{0}=0$, of the operator $\mathcal{L}$. All the other eigenvalues have negative real parts. The eigenvalue with the smallest nonzero real part, $e_{1}$, sets the longest time scale of the system. We have always found that $e_{1}$ is real, and defined $\tau_{r}=|e_{1}^{-1}|$. In order to determine this eigenvalue, $\mathcal{L}$ has been discretized and the eigenvalue problem $\mathcal{L}v = \lambda v$ solved, using suitable algorithms for sparse matrices. Stability against the discretization of the operator has been checked.
Figure~\ref{fig:fig4} shows the value of $\tau_{r}$ as a function of the bias $u$, in both the on-resonance and off-resonance cases. In both cases, $\tau_{r}$ decreases as $u$ is increased, and eventually saturates. Notice that the threshold value of $u$ for this saturation roughly corresponds to the rightmost fold bifurcation(s) in Fig.~\ref{fig:fig2}, i.e. to the entrance in the mono-stable region.\\
\noindent To determine $\tau_{d}$, a statistical analysis of several different solutions of the Langevin equation for different initial conditions and realizations of the noise, has been carried out for times up to $\tau=10^9$.
\begin{figure}[htb!]
 \centering
 \includegraphics[width=\columnwidth,keepaspectratio]{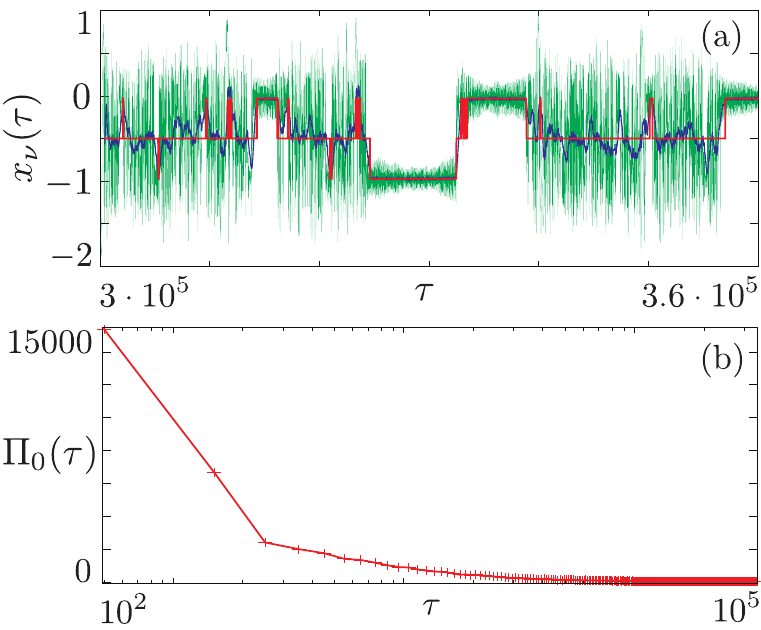}
 \caption{\label{fig:fig5} (a) A typical solution trace $x_{\nu}(\tau)$ (green curve), the running-averaged version (blue curve) and trigger-detection of the occupancy of oscillator minima (red curve). (b) Probability $\Pi_{0}(\tau)$ that the oscillator has spent a time $\tau$ in the minimum of $U(x)$ around $x=0$. Parameters: $u=0.2$, $\varepsilon=-0.5$, $\omega=10^{-3}$ and $\gamma=0.08$.}
\end{figure}
For a given realization of the noise $\xi_{\nu}(\tau)$, the corresponding solution $x_\nu(\tau)$ is smoothed by a running-average method. Subsequently, the regions where $|x_{\nu}-x_{0}^{(w)}|<\delta$, $|x_{\nu}-x_{1}^{(w)}|<\delta$ and $|x_{\nu}-x_{1/2}^{(w)}|<\delta$ are identified (trigger detection), where $x_{k}^{(w)}$ is the location of the minimum of the potential well corresponding to an average dot occupation $k$ ($k=0,1,1/2$), and $\delta$ is a threshold, set to 0.03 in our numerical analyses. This allows to determine the time $\tau_{k}^{(i)}$ spent during the $i$-th visit of well $k$. From a statistical analysis of the times $\tau_{k}^{(i)}$, the probability distribution $\Pi_{k}(\tau)$ that the oscillator has spent a given time $\tau$ in the $k$-th well is obtained; the procedure is then iterated over several different realizations of the noise process. The procedure is exemplified in Fig.~\ref{fig:fig5}(a) and the resulting probability distribution for the well centered at $x_{0}^{(w)}$ is shown in Fig.~\ref{fig:fig5}(b). Finally, a dwell time $\tau_{d}^{(k)}=\int{\mathrm d}\tau\,\tau\ \Pi_{k}(\tau)$ is evaluated. In the on-resonance case, one finds $\tau_{d}^{(0)}=\tau_{d}^{(1)}\equiv\tau_{d}$, whereas off-resonance two different dwell times are obtained.\\

\begin{figure}[htb!]
 \centering
 \includegraphics[width=\columnwidth,keepaspectratio]{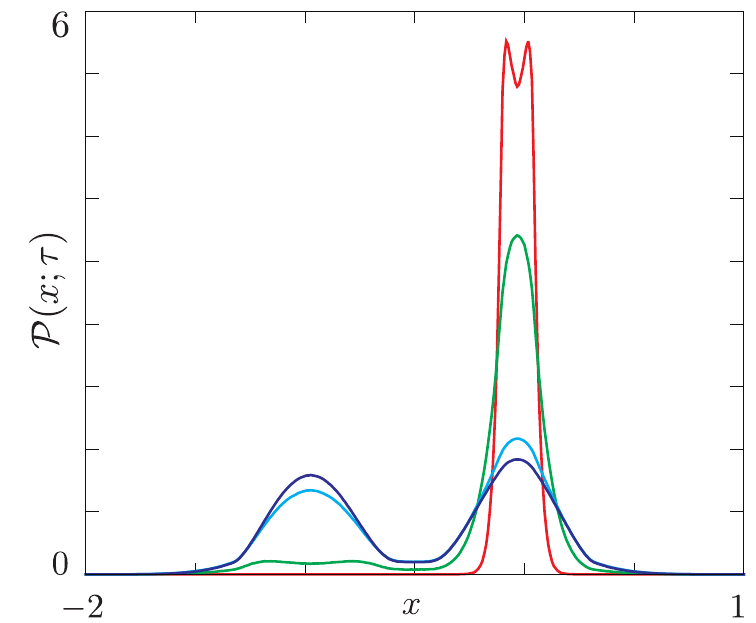}
 \caption{\label{fig:fig6} Reduced probability density $\mathcal{P}(x;\tau)$ as a function of $x$ for $u=0.2$, $\varepsilon=-0.5$ and different values of $\tau$, given  the initial condition $x_0=0, v_0=0$: $\tau=500$ (red); $\tau=2\cdot10^3$ (green); $\tau=5\cdot 10^{4}$ (cyan); $\tau=10^{6}$ (blue). Other parameters: $\omega=10^{-3}$, $\gamma=0.08$.}
\end{figure}
\noindent The orange and yellow vertical lines in Fig.~\ref{fig:fig3} denote the estimated values for the dwell and relaxation times $\tau_{d}\approx2\cdot 10^{3}$ and $\tau_{r}\approx4\cdot10^{4}$, respectively.
Figure~\ref{fig:fig6} shows snapshots at different times of the reduced probability density $\mathcal{P}(x;\tau)$ of the oscillator with initial condition $x_0=0, v_0=0$, one of the two cases shown in Fig.~\ref{fig:fig3}. For $\tau<\tau_{d}$ the occupation of the well about $x=-1$ is clearly negligible, and a stable steady state is reached only for $\tau>\tau_{r}$ (cyan and blue curves). Notice that the steady state density $\mathcal{P}(x;\tau)$ is symmetric with respect to $x=-0.5$, as implied by the symmetry of $U(x)$, $A(x)$, and $D(x)$ for $\varepsilon=-0.5$.\\
\noindent Figure~\ref{fig:fig5}(a) also proves our statement about the dynamical nature of the steady state: even in the long time regime, the oscillator position stochastically jumps between the minima of the effective potential and only the probability distribution satisfied $\dot{\mathcal{P}}(x,v;\tau)$.

\begin{figure}[htb!]
 \centering
 \includegraphics[width=\columnwidth,keepaspectratio]{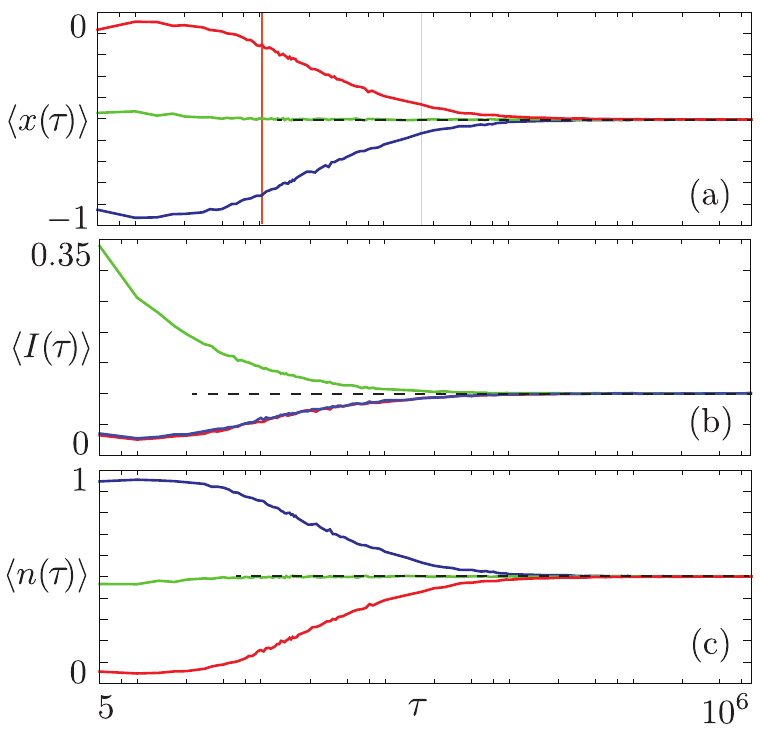}
 \caption{\label{fig:fig7} (a) $\langle x(\tau)\rangle$; (b) $\langle I(\tau)\rangle$; (c) $\langle n(\tau)\rangle$ for $u=0.475$, $\varepsilon=-0.5$ and different initial conditions $x_0=0, v_0=0$ (red curve), $x_0=-1, v_0=0$ (blue curve), and $x_0=-0.45, v_0=0$ (green curve). The orange (yellow) vertical line marks the characteristic dwell (relaxation) time scale $\tau_{d}$ ($\tau_{r}$). The steady state values are denoted by a dashed line. Other parameters: $\omega=10^{-3}$, $\gamma=0.08$.}
\end{figure}

\noindent The results in the tri-stable case are shown in Fig.~\ref{fig:fig7}, when the system is initialized near each of the three minima of $U(x)$. The behavior when starting at rest from $x_0=0$ and $x_0=-1$ (red and blue curves) is qualitatively similar to that for the bi-stable case, with shorter $\tau_{d}$ and $\tau_{r}$. The solution when starting near the third, middle, well is different (green curve). Indeed, both average position and dot occupation vary only slightly, whereas the current exhibits a marked decrease in time. No dwell time can be detected for this case, since, due to the noise term $D(x)$, the state quickly escapes from the central valley of $U(x)$ towards one of the lateral minima, see Fig~\ref{fig:fig1}(b).\\
\noindent The current traces are identical when the minima near $x=0$ or $x=-1$ are initially populated, similarly to the bi-stable case. The initial population of the central potential well (green curve) makes the current maximal, as discussed in Sec.~\ref{subsec:subsec2}. The current decreases as far as the lateral wells get occupied. We note that the decay time of $\langle I(\tau)\rangle$ roughly corresponds to $\tau_{r}$, which shows that for this kind of initial conditions the only time scale relevant to define the transient towards steady state is the relaxation time.\\
\begin{figure}[htb!]
 \centering
 \includegraphics[width=\columnwidth,keepaspectratio]{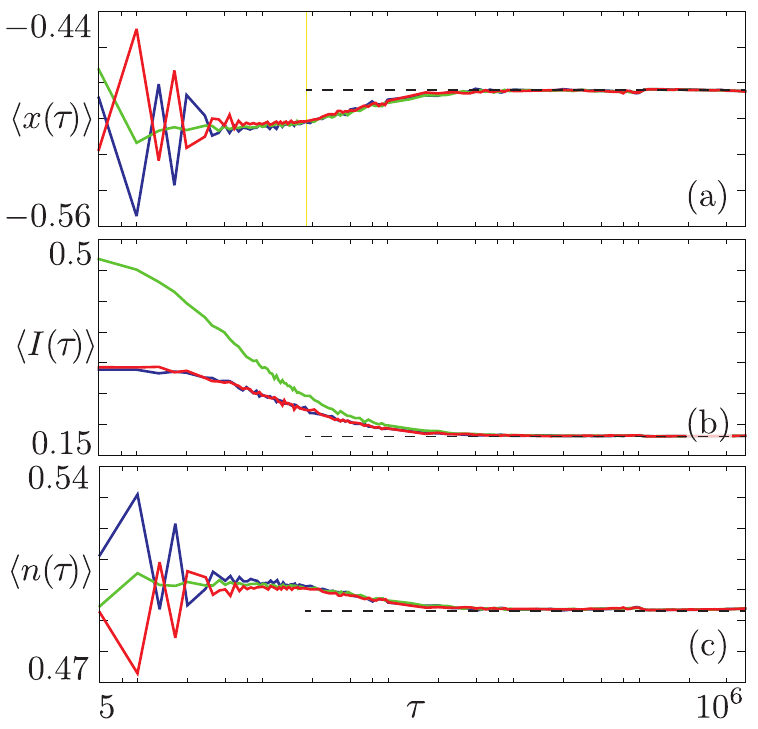}
 \caption{\label{fig:fig8} (a) $\langle x(\tau)\rangle$; (b) $\langle I(\tau)\rangle$; (c) $\langle n(\tau)\rangle$ for $u=1$, $\varepsilon=-0.5$ and different initial conditions $x_0=0, v_0=0$ (red curve), $x_0=-1, v_0=0$ (blue curve) and $x_0=-0.45, v_0=0$ (green curve). The yellow vertical line marks the characteristic relaxation time $\tau_{r}$. The steady state values are denoted by a dashed line. Other parameters: $\omega=10^{-3}$, $\gamma=0.08$.}
\end{figure}

\noindent For $u\geq\sqrt{2\gamma/\pi}$, only the minimum at $x=-0.5$ survives. Figure~\ref{fig:fig8} shows the average position, current and dot occupation versus $\tau$: in this case no dwell time can be defined as the dynamics is entirely restricted to the single central potential well. Indeed, all the solutions for the three initial conditions display essentially the same qualitative behavior and the calculated value of $\tau_{r}$ roughly matches the time when the transition towards the steady state occurs. Only the current trace for the initial condition $x_{0}=-0.5, v_{0}=0$ attains larger values at short times, since in this case the probability distribution is initially concentrated near the region where current is maximal. The oscillations shown in panels (a) and (c) are due to the small number of points considered on the time axis, which does not allow to sample in full details the short-time oscillations already mentioned above.

\subsubsection{Off-resonance case ($\varepsilon = -0.4$)}
\begin{figure}[htb!]
 \centering
 \includegraphics[width=\columnwidth,keepaspectratio]{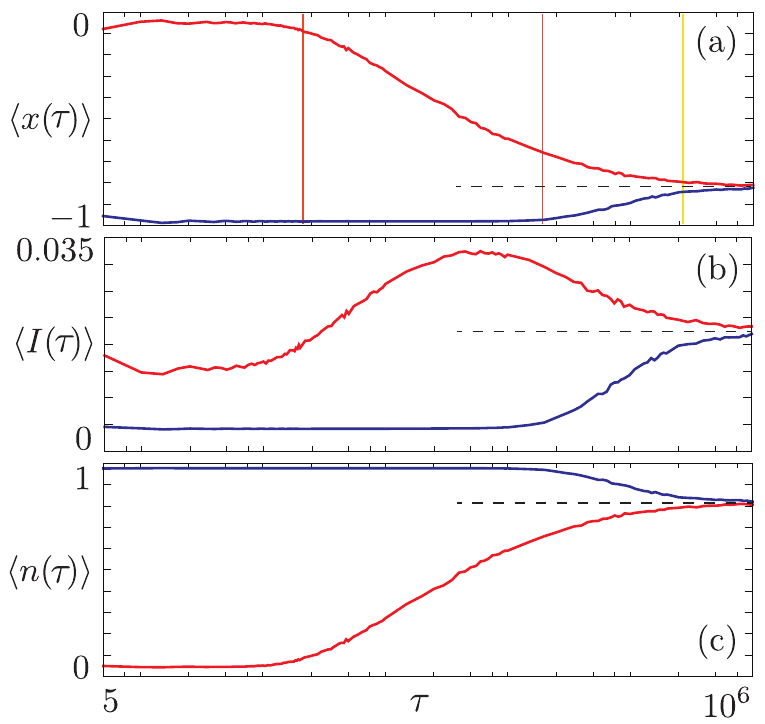}
 \caption{\label{fig:fig9} (a) $\langle x(\tau)\rangle$; (b) $\langle I(\tau)\rangle$; (c) $\langle n(\tau)\rangle$ for $u=0.2$, $\varepsilon=-0.4$ and different initial conditions $x_0=0, v_0=0$ (red curve), $x_0=-1, v_0=0$ (blue curve) and $x_0=-0.45, v_0=0$ (green curve). The orange (yellow) vertical line marks the characteristic dwell (relaxation) time scale $\tau_{d}$ ($\tau_{r}$). The steady state values are denoted by a dashed line. Other parameters: $\omega=10^{-3}$, $\gamma=0.08$.}
\end{figure}
\noindent Figure~\ref{fig:fig9} shows the case $u=0.2$ (see Fig.~\ref{fig:fig1}(d)). The most striking feature, as already anticipated in Sec.~\ref{subsec:subsec2}, is the presence of two different dwell times corresponding to the two different potential wells, with the most stable one having the largest $\tau_{d}$, as expected. The average steady position decreases towards $x=-1$ and, correspondingly, the occupation of the dot tends to a value larger than $0.5$, since in this case the steady state cannot be expected to be symmetrical with respect to $x=-0.5$. More intriguing is the behavior of the current as a function of time, when the oscillator is initialized around $x=0$, namely in the least favorable well. Here, the current exhibits a non-monotonous behavior with a sharp increase leading to a maximum for $\tau=\tau^{*}\approx 5500$, followed by a decrease towards the asymptotic value.
\begin{figure}[htb!]
 \centering
 \includegraphics[width=\columnwidth,keepaspectratio]{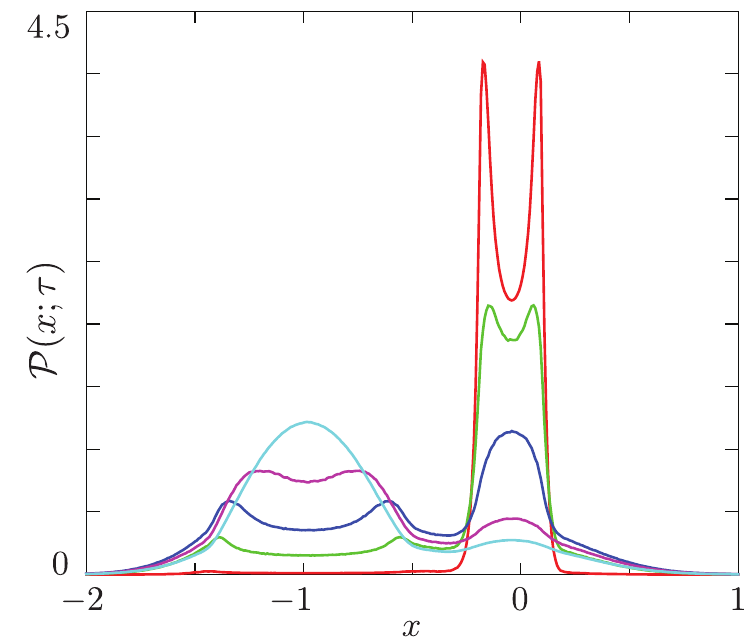}
 \caption{\label{fig:fig10} Reduced probability density $\mathcal{P}(x;\tau)$ for $u=0.2$, $\varepsilon=-0.4$ and different values of $\tau$, given  the initial condition $x_0=0, v_0=0$: $\tau=150$ (red); $\tau=1000$ (green); $\tau=5.5\cdot 10^{3}$ (blue); $\tau=10^{4}$ (purple); $\tau=4\cdot10^5$ (cyan). Other parameters: $\omega=10^{-3}$, $\gamma=0.08$.}
\end{figure}
In order to understand this behavior, in Fig.~\ref{fig:fig10} we show the reduced probability density $\mathcal{P}(x;\tau)$ for different values of $\tau$ starting from the initial condition $x_0=0, v_0=0$. As $\tau$ approaches $\tau^{*}$, the probability density near $x\approx-0.4$ increases monotonically. For $\tau>\tau^{*}$ the trend is reversed. Furthermore, it can be checked using Eq.~(\ref{eq:I}) that for $\varepsilon=-0.4$, $I(x)$ attains its maximum around $x\approx-0.4$. It is worth stressing again that even in this case, due to the extremely long time scales for small bias $u$, an observation up to $\tau_{d}$ would not show the reaching of a common steady state for different initial conditions of the oscillator and may even lead to an apparent divergence of the $\langle I(\tau)\rangle$ traces.
\subsection{Discussion}
\subsubsection{Validity of the model and of the strong coupling regime}
\label{sec:validity}
The main assumption of this paper is the adiabatic condition
$\omega\ll\gamma$, which allows to neglect the dot transient dynamics
in Eqns.~(\ref{eq:Ag},\ref{eq:Dg},\ref{eq:Fg}) and ensuing
Eqns.~(\ref{eq:A},\ref{eq:D},\ref{eq:F}) which only retain the
oscillator time dependence $x(t)$. Indeed, this
approximation remains valid as long as the adiabatic condition is
fulfilled and one explores time scales slower
$\omega^{-1}$. Increasing $\omega\sim\gamma$ breaks the validity of
the above approximation~\cite{brandes2}. In this case, it can be
expected that qualitative modifications can occur with respect to the
short-time behavior described in this paper. However, such
modifications should not spoil the uniqueness of the steady
state. This interesting topic is beyond the scope of the present paper
and may be the subject of future investigations.\\
\noindent {Additionally, in order to observe bistability and tristability, the strong coupling condition $\gamma<1$, $\omega\ll1$ must be fulfilled~\cite{pistolesi}, as mentioned in Sec.~\ref{sec:sec1}.}\\

\noindent Quantum fluctuations beyond the damping and noise terms in
the semiclassical Langevin equation may in principle be
present~\cite{mitra,dykman}. Such fluctuations may as well, influcence
the transient dynamics of the system affecting the hopping probability
of the oscillator state. However, they have been shown to be relevant
when $u<\omega$~\cite{mozyrsky,pistolesi,nocera}. Thus, all results
presented here are well within the domain of the semiclassical
approximation.\\

\noindent Thermal flutuations induced by an extrinsic thermal
bath as well as intrinsic mechanical damping mechanisms have been
neglected in this paper. They can be modeled by an additional
(constant) damping term in $A(x)\to A(x)+\eta$ and diffusion term
$D(x)\to D(x)+\eta T_\eta$ where $T_{\eta}=k_{B}T_{b}/2E_{p}$ and
$T_{b}$ is the effective bath temperature~\cite{pistolesi}. Both the
enhancement of damping and hoppping are expected to quantitatively
affect the time scales discussed in the paper. In particular, thermal
fluctuations are expected to lead, in the regime of strong coupling to
this additional effective thermal bath, to an enhancement of the
hopping rate and to a shortening of the dwell time $\tau_{d}$. {Physically, we can expect that our results are not qualitatively affected if $\eta< A_{\mathrm{max}}$ and $\eta T_{\eta}< D_{\mathrm{max}}$ where we can estimate $A_{\mathrm{max}}\approx\omega/\gamma^2$ and $D_{\mathrm{max}\approx\omega/\gamma}$. For the parameters employed in this paper, one gets $A_{\mathrm{max}}\approx10^{-1}$ and $D_{\mathrm{max}}\approx10^{-2}$.\\ 
\noindent Let us now relate these conditions to experimental parameters. Supposing that for $\eta=\omega/\gamma^{2}=A_{\mathrm{max}}$ the damping is essentially dominated by the extrinsic bath and still assuming a weak renormalization of the system frequency due to damping, one can identify the quality factor $Q\approx\eta^{-1}$. Thus, the condition $\eta< A_{\mathrm{max}}$ is equivalent to $Q\gg A_{\mathrm{max}}^{-1}$. For the case considered in this paper, $Q>10$ is therefore required. As will be seen, systems such as suspended CNTs can easily exceed this limitation.\\
\noindent Assuming the worst-case scenario $\eta=\omega/\gamma^2$, the condition $\eta T_{\eta}< D_{\mathrm{max}}$ can be cast into a bound for $T_{\eta}$ which implies $k_{B}T_{b}<\Gamma_{0}$. Assuming a typical $\Gamma_{0}\approx10\ \mathrm{GHz}$ one has $T_{b}<0.1\ \mathrm{K}$.}
\subsubsection{Observability of the proposed effect}
\label{sec:obs}
\noindent The fact that the oscillator properties ($\langle x(\tau)\rangle$) are directly related
to those of the dot ($\langle I(\tau)\rangle$) make our results
particularly appealing, since the latter should in principle be
detectable much more easily in a transport experiment.\\
\noindent {As already discussed above, in order to detect the dwell and relaxation time scales
$\tau_{d}$ and $\tau_{r}$, one needs to consider a strong coupling between the electrons and the adiabatic oscillator. In addition to the requirements about the quality factor and extrinsic thermal bath temperatures, further restrictions apply on the bare system oscillator frequency, tunneling rate and the electron-vibration coupling force. In the following we will review three of the most appealing candidates to observe the transient dynamics, namely nano-beams~\cite{blencowe}, suspended carbon nanotubes~\cite{charlier} and molecular systems~\cite{molsys1,molsys2}}\\
{Nano-mechanical cantilevers, such as those that may be created in
Si structures~\cite{blencowe} are characterized by typical bare vibrational frequencies as low as 1$\div$10
MHz, which would yield $\tau_{d}\approx1\div10\ \mathrm{ms}$ and
$\tau_{r}\approx10\div100\ \mathrm{ms}$ at small bias. Similar performances
may be obtained with metal nano-beams coated with a semiconducting
piezoresistive layer~\cite{tang}. With such low vibrational frequencies, satisfying the adiabatic condition poses no particular challenge. The most relevant coupling mechanism between electrons and the oscillations is due to electrostatic gating. The capacitance of the beam with respect to an external gate fluctuates as the flexural mode is excited, leading to a bilinear coupling between the excess charge on the nano-beam and the amplitude of the flexural mode. Estimates show that such coupling is not very large~\cite{blencowe}, although it can be in principle tuned by a suitable a gate voltage.}\\
\noindent Another possible candidate is a suspended carbon
nanotube~\cite{charlier} {considering the lowest flexural mode. Such a system can easily satisfy the adiabatic condition and exhibit strong electron-vibron coupling. In view of the great interest and relevance of this system, we provide here actual estimates based on state-of-the-art experimental setups. We begin discussing the coupling between electrons and vibrations, which occurs in two different ways. The {\em intrinsic} electron-phonon mechanism~\cite{eroscoup} would lead to a coupling term quadratic in the displacement operator $X$. This coupling, however, has been shown to be very weak~\cite{eroscoup} and unlikely to give rise to bistability. However, also in this case an external biased gate can be employed to couple the flexural mode and electrons. In this case, the coupling is {\em linear} in $X$ and has the form of Eq.~(\ref{eq:hdv}) with the electron-phonon coupling force given by~\cite{armcoup}
\begin{equation}
\lambda=\frac{|e|C_{g}V_{g}}{(C_{0}+C_{g})\ln\left(d/r\right)d}\, ,	
\end{equation}
where
\begin{equation}
C_{g}=\frac{2\pi\varepsilon_{0}L}{\ln\left(d/r\right)}	
\end{equation}
is the CNT-gate capacitance, with $\varepsilon_{0}$ the vacuum permittivity, $L$ the CNT length, $d$ the distance between the CNT and the gate and $r$ the CNT radius. Finally, $C_{0}$ is the capacitance of the CNT with respect to all other gates, and $V_{g}$ the voltage present on the gate.\\
\noindent We will consider a CNT with typical parameters $L\approx1.8\ \mu\mathrm{m}$, $r\approx1.5\ \mathrm{nm}$, $d\approx400\ \mathrm{nm}$ and a typical gate voltage $V_{g}\approx 1\ \mathrm{V}$~\cite{prlexp}. We also assume a typical value of $C_{0}\approx10^{-17}\ \mathrm{F}$~\cite{prbcap}. One obtains $\Omega_{0}\approx200\ \mathrm{MHz}\approx0.8\ \mu\mathrm{eV}$ where a stiffening contribution due to the additional strain induced by the gate has been accounted for~\cite{footnote}, $C_{g}\approx 2\cdot10^{-17}\ \mathrm{F}$ and thus $E_{p}=\lambda^2/2m\Omega_{0}^{2}\approx15\ \mu\mathrm{eV}$, with $m\approx10^{-20}\ \mathrm{Kg}$. This implies $\omega\approx 0.02\ll 1$. Setting $\gamma=10\omega=0.2<1$ implies a typical tunneling rate $\Gamma_{0}\approx6\ \mathrm{GHz}$ corresponding to an average current $I_{0}\approx1\ \mathrm{nA}$, well within the range of experiments~\cite{prlexp}.} With {the above estimates we get} $\tau_{d}\approx 0.1\ \mathrm{ms}$
and $\tau_{r}\approx\ 1\mathrm{ms}$. Extremely high quality factors
$Q\gtrsim 10^5$ have been reported for the flexural mode of suspended
CNTs~\cite{highQ}, constituting a favorable condition to observe the
predicted effects. Other vibrational modes or tensile-stressed CNTs~\cite{cobdenhi} have
frequencies at least two orders of magnitude larger~\cite{sapmaz},
making the detection of the above times more problematic.\\
\noindent {Finally, also molecular systems exhibit a strong coupling between electrons and vibrations~\cite{pistolesi}. However, the coupling to the leads is generally very weak, which in addition to the usually large oscillation frequencies makes it extremely difficult to achieve the adiabatic regime~\cite{pistolesi,pistolaba} considered in this paper.}
\section{Conclusions}
\label{sec:sec3}
In this paper we have investigated the time evolution towards the
steady state of a NEMS in the adiabatic regime. The Langevin equation
for the oscillator and the corresponding Fokker-Planck equation for
the probability density in phase space are numerically solved for
arbitrary time and initial conditions in several parameter settings of
the system, corresponding to bi-stability (low bias), tri-stability
(intermediate bias and resonance) or mono-stability (large bias). It
is shown that in all situations a unique steady state is reached. The
approach to the steady state is however different according to the
considered parameter setting. For low bias, the dynamics of the system
is found to be essentially frozen in a sort of blockaded regime up to
rather large time scales, before relaxing towards the steady
state. Two time scales emerge, corresponding to the dwell time into
quasi-stable minima of the effective potential well of the oscillator
and to the relaxation time towards the steady situation. For
intermediate bias, the dynamics is described by either one or two time
scales, depending on the initial condition of the system. For large
bias, the system relaxes towards the steady state with a single time
scale. Dwell and relaxation time scales have been identified
numerically by analyzing the statistical properties of the solutions
of the Langevin equation and the spectral properties of the
Fokker-Planck operator. Our findings suggest that no hysteretic
behavior is shown by this system at steady state, but that an
observation of the system dynamics stopping at short time scales might
fail to exhibit the tendency towards a unique steady state, especially
at small bias and/or off-resonance~\cite{albrecht}. The predicted time
scales are within the range of current experiments.\\

\noindent The support of MIUR via MIUR-FIRB2012 - Project HybridNanoDev, Grant
No. RBFR1236VV, and of University of Genoa are gratefully acknowledged.


\begin{thebibliography}{99}
\bibitem{footnote} {The tension induced by the pulling gate gives rise to a small nonlinear term $\propto X^{3}$ in in the effective potential of the oscillator. We have checked that this term is negligible for the regimes discussed in this paper}.
\end{thebibliography}
\end{document}